\documentstyle[buckow]{article}

\newcommand{\bra}[1]{\langle{#1}|}
\newcommand{\ket}[1]{|{#1}\rangle}
 
\begin {document}
\rightline{DFTT 49/2000}
\def\titleline{
D-branes in type I string theory
\footnote{Talk given by L.~Gallot at the workshop 
``The quantum structure of spacetime and the geometric 
nature of fundamental interactions'' in Berlin, October 2000}
}
\def\authors{
M. Frau\1ad, L. Gallot\1ad, A. Lerda\1ad\2ad and 
P. Strigazzi\1ad
}
\def\addresses{
\1ad Dipartimento di Fisica Teorica, Universit\`a di Torino and
I.N.F.N., Sezione di Torino, \\
Via P. Giuria 1, I-10125 Torino, Italy\\
\2ad Dipartimento di Scienze e Tecnologie Avanzate, Universit\`a del
Piemonte Orientale, \\
Corso G. Borsalino 54, I-15100 Alessandria, Italy
}
\def\abstracttext{
We review the boundary state description of
D-branes in type I string theory and show that the only stable 
non BPS configurations are the D-particle and the D-instanton.
We also compute the gauge and gravitational interactions of the non-BPS 
D-particles and compare them with the interactions of the 
dual non-BPS particles of the heterotic string, 
finding complete agreement.
}
\large
\makefront
\section{D-branes of type II theories}
The Dirichlet branes \cite{pol}
can be described, in the weak coupling regime of string theory, by a boundary 
conformal field theory and admit a two-fold interpretation: 
on the one hand, they are objects on which open 
strings can end, and on the other hand they  
can emit or absorb closed strings.
Therefore, introducing D-branes in a theory of closed strings amounts 
to extend their conformal field theory by introducing
world-sheets with boundaries and imposing appropriate boundary
conditions on the closed string coordinates $X^\mu$. In the operator
formalism these boundary conditions are implemented through
the so called boundary state $\ket{Dp}$ \cite{pa},
whose bosonic part is defined by the following eigenvalue 
problem
\begin{equation}
\partial_{\tau}X^{\alpha}(\sigma , 0)\, \ket{Dp}_X =0~~,~~
\Big(X^{i}(\sigma , 0)-x^i\Big) \ket{Dp}_X = 0~~,
\label{boundcond}
\end{equation}
where $\alpha=0,\ldots,p$ labels the longitudinal
directions, $i=p+1,\ldots,9$ labels the transverse
directions and the $x^i$'s denote the position of the brane in the 
transverse space.
World-sheet supersymmetry requires that analogous equations must
be also imposed on the left and right moving fermionic fields 
$\psi^\mu$ and ${\widetilde \psi}^\mu$, thus defining the
fermionic part of the boundary state
\begin{equation}
\Big(\psi^\alpha(\sigma,0) - {\rm i}\,\eta\,{\widetilde \psi}^\alpha(\sigma,0)
\Big)\ket{Dp,\eta}_\psi = 0 ~~,~~
\Big(\psi^i(\sigma,0) +{\rm i}\,\eta\,{\widetilde \psi}^i(\sigma,0)
\Big)\ket{Dp,\eta}_\psi = 0 ~~,
\label{boundcond1} 
\end{equation}
where $\eta=\pm1$. Notice that there are two consistent implementations
of the fermionic  boundary conditions corresponding to the
sign of $\eta$, and consequently there are two different boundary states
\begin{equation}
\ket{Dp,\eta} = \ket{Dp}_X\,\ket{Dp,\eta}_\psi
\label{bs}
\end{equation}
both in the NS-NS and in the R-R sectors.
The overlap equations (\ref{boundcond}) and (\ref{boundcond1})
allow to determine the explicit structure of the boundary
states (\ref{bs}) up to an overall factor. This normalization
can then be uniquely fixed by factorizing amplitudes with 
closed strings emitted from a disk and turns out
to be given by (one half of) the brane tension measured in
units of the gravitational coupling constant, {\it i.e.} 
$T_p=\sqrt{\pi}\,\Big(2\pi\sqrt{\alpha'}\Big)^{3-p}~$.
We would like to remark that even if each boundary state
$\ket{Dp,\eta}$ is perfectly consistent from the conformal
field theory point of view, not all of them are
acceptable in string theory. In fact, to describe a 
physical D-brane a boundary state has to
satisfy three requirements \cite{BerGab}:
 
{\it i}) to be invariant under the closed string GSO projection (and 
also under orbifold or orientifold projections if needed); 

{\it ii}) the tree level amplitude due to the exchange of 
closed strings between two boundary states, after modular transformation, 
has to make sense as a consistent open string 
partition function at one-loop;

{\it iii}) the open strings introduced through the D-branes must have 
consistent couplings with the original closed strings and eventually, 
have to be compatible with the orbifold/orientifold projection.

Using these prescriptions, it is rather simple
to find the boundary state for the supersymmetric BPS D$p$-branes 
of type II.
In particular, the GSO projection of the type II theories 
forces us to retain only the following linear combinations
\begin{equation}
\ket{Dp}_{\rm NS} = 
\frac{1}{2} \Big[ \ket{Dp,+}_{\rm NS} 
- \ket{Dp,-}_{\rm NS} \Big] ~~,~~
\ket{Dp}_{\rm R} =
\frac{1}{2} \Big[ \ket{Dp,+}_{\rm R} 
+\ket{Dp,-}_{\rm R} \Big]
\label{bsgso}
\end{equation}   
in the NS-NS and in the R-R sectors respectively, with $p=0,2,4,6,8$ for IIA, 
$p=-1,1,3,5,7,9$ for IIB. 
To read the spectrum of the open strings living on the D$p$-brane 
(called $p$-$p$ strings),
one has first to evaluate 
the closed string exchange amplitude
$\bra{Dp}\,P\,\ket{Dp}~$
where $P$ is the closed string propagator, and then perform a modular
transformation to the open string
channel. Applying this procedure, one finds 
the following relations
\begin{eqnarray}
{}_{\rm NS}\bra{Dp,\eta}\,P\,\ket{Dp,\eta}_{\rm NS} 
&=& \int_0^\infty\frac{ds}{s}\,{\rm Tr}_{\rm NS} \,q^{2L_0-1}~~,\nonumber \\ 
{}_{\rm NS}\bra{Dp,\eta}\,P\,\ket{Dp,-\eta}_{\rm NS} &=&
\int_0^\infty\frac{ds}{s}\,{\rm Tr}_{\rm R}\, q^{2L_0} ~~,\nonumber \\
{}_{\rm R}\bra{Dp,\eta}\,P\,\ket{Dp,\eta}_{\rm R}
&=&\int_0^\infty\frac{ds}{s}\,{\rm Tr}_{\rm NS}\,(-1)^F\,
 q^{2L_0-1} ~~,\nonumber \\
{}_{\rm R}\bra{Dp,\eta}\,P\,\ket{Dp,-\eta}_{\rm R}          
&=&\int_0^\infty\frac{ds}{s}\,{\rm Tr}_{\rm R}\,(-1)^F\,
 q^{2L_0} = 0~~,
\label{openclosed}
\end{eqnarray} 
where $q={\rm e}^{-\pi s}$. It is then clear that in order to
obtain the
supersymmetric ({\it i.e.} GSO projected) open string amplitude 
\begin{equation}
{\cal Z}_{open,BPS}^{p-p}\int_0^\infty\frac{ds}{s}\left[{\rm Tr}_{\rm NS}\left(\frac{1+(-1)^F}{2}
\right)q^{2L_0-1} 
- {\rm Tr}_{\rm R}\left(\frac{1+(-1)^F}{2}
\right)q^{2L_0} \right]~~,
\end{equation}
one must consider the following boundary state
\begin{equation}
\ket{Dp} = \ket{Dp}_{\rm NS} \pm \ket{Dp}_{\rm R}
\label{bsbps}
\end{equation}
where the sign ambiguity is related to the existence of branes and anti-branes.

The criteria $i)$ - $iii)$
defining physical D-branes do not rely 
at all on {\it space-time} supersymmetry, and thus one may wonder whether 
in type II theories there may exist also non-supersymmetric branes. 
This problem has been systematically addressed in a series of
papers by A. Sen \cite{sen}, who constructed 
explicit examples of non-BPS  branes.
In particular, he considered the superposition of a 
D-string of type IIB and an anti-D-string (with a $Z_2$ Wilson line) 
and by suitably 
condensing the tachyons of the open strings stretching between the brane
and the anti-brane, he managed to construct a new configuration of type IIB 
which behaves like a D-particle, does not couple to any R-R field 
and is heavier by a factor of $\sqrt{2}$ than the
BPS D-particle of the IIA theory. The boundary state for this non-BPS 
D-particle has been explicitlely constructed in Ref.~\cite{gallot}.
This 
construction can be obviously generalized to the case 
of a pair formed by two BPS D$(p+1)$-branes with opposite R-R charge
(and with a $Z_2$ Wilson line) which, after tachyon condensation, 
becomes a non-BPS D$p$-brane. 
Alternatively, this same non-BPS
configuration can be described starting from a
superposition of two BPS D$p$ branes with opposite R-R charge 
and modding out the theory by the operator $(-1)^{F_L}$ whose effect is to 
change the sign of all states in the R-R and R-NS sectors. In this second
scheme, a superposition of a D$p$-brane and anti-D$p$-brane of type IIA (IIB)
becomes in the reduced theory a non-BPS D$p$-brane of type IIB (IIA).
In either way we therefore find that there
exist non-BPS D$p$-branes for $p = -1,1,3,5,7,9$ in IIA and $p=0,2,4,6,8$ IIB.
For reviews on this subject, see Refs.~\cite{reviews,BerGab}.

A boundary state interpretation can be given to these non BPS branes
provided they satisfy conditions ${\it i})$ - ${\it iii})$ mentioned above.
Being manifestly non supersymmetric, the spectrum of open strings on their 
world volume is itself non supersymmetric. The corresponding vacuum amplitude 
in the open channel is in fact given by
\begin{equation}
{\cal Z}_{open, non BPS}^{p-p}=
\int_0^\infty\frac{ds}{s}\left[{\rm Tr}_{\rm NS}q^{2L_0-1} 
-{\rm Tr}_{\rm R}q^{2L_0} \right]~~,
\label{ns}
\end{equation}
where there is no GSO projection on the open string spectrum. Condition 
{\it ii}) and relations (\ref{openclosed}) then 
imply that the corresponding boundary state is, for any value of $p$,
\begin{equation}
\ket{Dp} = 
\sqrt{2} \,
\ket{Dp}_{\rm NS}~~,
\label{tens1}
\end{equation} 
thus confirming the facts that non BPS branes are heavier by a factor of 
$\sqrt{2}$ than the BPS branes of same dimension and are neutral under R-R 
forms. Moreover, for a given theory, not any value of $p$ is acceptable. 
Indeed the closed string amplitude between a BPS and a non BPS brane of same 
dimension reads
\begin{equation}
_{BPS}\bra{Dp} P \ket{Dp}_{nonBPS} =
\frac{1}{\sqrt{2}}\int_0^\infty\frac{ds}{s}\left[{\rm Tr}_{\rm NS}
\,q^{2L_0-1} 
-{\rm Tr}_{\rm R}\,q^{2L_0} \right]~~,
\end{equation}
which is not a sensible open string partition function due to the global 
irrational factor $1/\sqrt{2}$. Thus, whenever a BPS Dp-brane exists, 
a non BPS one cannot, hence we conclude that non BPS branes exist for 
$p = -1,1,3,5,7,9$ in IIA and $p=0,2,4,6,8$ IIB.

It is clear from (\ref{ns}) that the non-BPS branes of type II
are not stable, because the absence of the GSO projection on the open strings
leaves the NS tachyon on their world-volume. 
However, they could become stable
in an orbifold of type II theory, say IIA(B)$/{\cal P}$, 
provided that the tachyon be odd under the projection ${\cal P}$. 
In the orbifold theory in fact, the non-BPS 
vacuum amplitude of the $p$-$p$ open-strings reads
\begin{equation}
{\cal Z}_{open}=
\int_0^\infty\frac{ds}{s}\left[{\rm Tr}_{\rm NS}\left(\frac{1+{\cal P}}{2}
\,q^{2L_0-1} \right)
-{\rm Tr}_{\rm R}\left(\frac{1+{\cal P}}{2}
\,q^{2L_0}\right) \right]~~,
\label{zop1}
\end{equation}
and the natural question to 
ask is to which boundary state it corresponds.
In the case of a space-time orbifold, the perturbative spectrum 
of the bulk theory contains only 
closed strings which can be untwisted or twisted under the 
orbifold. Therefore, there are several 
sectors to which the bosonic states belong, 
and there exist different types
of boundary states depending on which components in those sectors
they have. The twisted boundary states then account for the piece depending 
on ${\cal P}$ in (\ref{zop1}).
In the case of a world-sheet orbifold, however, this simple 
picture does not hold. To illustrate this point, we
shall consider in the next section the specific case of the type I theory.

\section{D-branes of type I theory}
The type I theory is the orbifold of the IIB theory by the world-sheet 
parity $\Omega$. 
The distinctive feature of this model is that 
the perturbative states of the
twisted sector of the {\it bulk} theory now 
correspond to unoriented {\it open} 
strings which should then be appropriately incorporated in the
boundary state formalism. Let us briefly summarize how
this is done \cite{ps}.
The starting point is the projection of the closed string spectrum 
onto states which are invariant under $\Omega$. 
The corresponding closed string partition function is
obtained by adding a Klein bottle 
contribution to the modular invariant (halved) torus contribution. 
The Klein bottle is a genus one non-orientable 
self-intersecting surface which may be seen equivalently
as a cylinder ending at two crosscaps. A crosscap is a line of 
non-orientability, a circle with opposite points identified, and thus the 
associated crosscap state $\ket{C}$ is defined by
$$
X^{\mu}(\sigma+\pi , 0)\,\ket{C}= X^{\mu}(\sigma , 0)\,\ket{C} ~~,~~
\partial_{\tau}X^{\mu}(\sigma+\pi , 0)\,\ket{C} = -\partial_{\tau}X^{\mu}
(\sigma , 0)\,\ket{C}~~,
$$
and by the analogous relations appropriate for world-sheet fermions. 
As is clear from these equations, the crosscap state is 
related to the boundary state of the BPS space-time filling D9 brane
through $\ket{C} ~ \propto ~{\rm i}^{L_0+\tilde{L}_0}\,\ket {D9}~$, and its
normalization turns out to be 32 times the normalization of 
the boundary state for the D9-brane. 
Consequently, the (negative) charge 
for the unphysical 10-form R-R potential created by the
crosscap must be compensated by
the introduction of 32 D9 branes. 
In this way we then introduce unoriented open strings 
starting and ending on these 32 D9 branes, whose
vacuum amplitude is given by
\begin{equation}
{\cal Z}_{open}^{9-9} =
{1 \over 2} \Big(2^{10} \bra{D9}P\ket{D9}
+2^5 \bra{D9}P\ket{C} + 2^5\bra{C}P\ket{D9}\Big)~~.
\label{ZopI99}
\end{equation}
By adding to (\ref{ZopI99}) 
the contribution of the Klein bottle we obtain 
an expression, in which the tadpoles 
for the massless unphysical states cancel. A 
moment thought shows that this corresponds 
to choose the open string gauge group to be $SO(32)$. 
Thus, we can say that the type I theory possesses a 
``background'' boundary state given by
\begin{equation}
{1 \over \sqrt{2}} \Big(\ket{C}+32\ket{D9} \Big)~~,
\label{background}
\end{equation}
where the factor of $1/\sqrt{2}$ has been introduced 
to obtain the right normalization of the various spectra.
Performing a modular transformation, we can rewrite
the amplitude of eq. (\ref{ZopI99})
in the open string channel
as follows
\begin{equation}
{\cal Z}_{open}^{9-9} =\int_0^\infty\frac{ds}{s}
\left[{\rm Tr}_{\rm NS}\left( {{1+(-1)^{F} \over 2}} 
{1+ \Omega \over 2 }\,q^{2L_0-1}\right)
-{\rm Tr}_{\rm R}\left( {{1+(-1)^{F} \over 2}} 
{1+ \Omega \over 2 }\,q^{2L_0}\right)\right]~~,
\end{equation}
where the part depending on $\Omega$ comes from the M\"obius contribution. 
Thus, we see that in the type I theory the crosscap state plays 
the same role that the twisted part of the boundary state had 
in the space-time orbifolds. This means that the vacuum amplitude 
for a generic D$p$-brane is given by the half-sum of a cylinder and M\"obius strip 
contribution as follows
\begin{equation}
{\cal Z}_{open}^{p-p} =
{1 \over 2} \Big(\bra{Dp}P\ket{Dp}
\label{ZopI}
+\bra{Dp}P\ket{C} + \bra{C}P\ket{Dp}\Big)~~.
\end{equation}
In addition, the presence of background D9-branes allows for open strings 
stretching between a D$p$ and one of the 32 D9-branes, whose partition 
function is given by the mixed amplitude 
\begin{equation}
{\cal Z}_{open}^{p-9 \oplus 9-p}
={32 \over 2} \Big(\bra{Dp}P\ket{D9} +\bra{D9}P\ket{Dp}\Big)~~.
\label{p-9}
\end{equation}
In the following, we shall use this remark in order classify and obtain 
the boundary states for BPS and stable non-BPS branes of type I theory.

We have seen before that the type IIB theory contains BPS D$p$-branes with 
$p=-1,1,3,5,7,9$. The BPS D-branes of type I are then BPS branes of type II 
even under the $\Omega$ projection, a fact which occurs for $p=1,5,9$, the R-R part 
of $\ket{Dp}$ being odd under $\Omega$ for the other values, $p=-1,3,7$. We 
thus have in principle D$p$-branes with the following boundary state
\begin{equation}
{1 \over \sqrt{2}}\ket{Dp} = {1 \over \sqrt{2}}\left(\ket{Dp}_{\rm NS} + \ket{Dp}_{\rm R}
\right) ,
\label{bpsI}\end{equation}
where the factor of $1/\sqrt{2}$ is there as in (\ref{background}) 
in order to obtain the right normalization of vacuum amplitudes. 
A moment thought shows that the $\Omega$ projection has the effect to reduce 
the world volume gauge group on N branes from $SU(N)$ to $SO(N)$, 
accordingly with the vacuum amplitude here obtained. However, there is a 
subtlety in the case of the D5-brane. Indeed, due to the particular form 
of the M\"obius strip in that case, it is not possible to interpret the 
vacuum amplitude obtained from a boundary state of the type (\ref{bpsI}) 
with $p=5$ as a sensible partition function of open strings, {\it e.g.} 
containing a gauge vector in the adjoint representation of some group. 
In fact, the D5-brane of type I is the projection of 2 D5-branes of type 
II so that its boundary state is   
\begin{equation}
{2 \over \sqrt{2}}\left(\ket{D5}_{\rm NS} + \ket{D5}_{\rm R}
\right) .
\label{bpsI5}\end{equation}
In particular the R-R density charge of the D5-brane is doubled so that 
the Dirac quantization rule relating the D1 and D5 R-R densities is still 
valid. From the vacuum amplitude we directly see that the gauge 
group on the world volume of N D5-branes is $Sp(2N)$.

The type IIB theory contains unstable
non-BPS D$p$-branes with $p=0,2,4,6,8$ which are described
by the boundary state (\ref{tens1}). Now, we address
the question whether these D-branes become stable in the
type I theory, {\it i.e.} we examine whether the tachyons
of the $p$-$p$ open strings are removed by $\Omega$. As
explained in \cite{gallot}, the world-sheet parity can be
used to project the spectrum of the $p$-$p$ strings
only if $p=0,4,8$. Indeed, only in these cases $\Omega^2=1$ on the 
$p$-$p$ open strings, which are thus compatible with the orbifold projection 
(condition {\it ii}). Thus, the non-BPS D2 and D6 branes
will not be further considered. However, 
we must take into account also another
kind of configuration, namely the superposition of
a BPS D$p$-brane and an anti-D$p$-brane of type IIB
which clearly does not carry any R-R charge. 
This configuration is unstable due to the presence
of tachyons in the open strings stretching between the
brane and the anti-brane, but in the type I theory
these tachyons might be projected out. A systematic
analysis \cite{WITTEN,gallot} shows that in this case
$\Omega$ can be used as a projection only if $p=-1,3,7$. 

In conclusion, we have to analyze the stability of the
non-BPS D$p$-branes of type I with $p=-1,0,3,4,7,8$
whose corresponding boundary states 
are given by
\begin{equation}
{\mu_{p}\over \sqrt{2}}\ket{Dp}_{NS}
\end{equation} with suitable positive values of
$\mu_p$.
To address this problem, we need to consider the spectrum
of the unoriented strings living on the
brane world-volume (the $p$-$p$ sector),
and also the spectrum of the open strings
stretched between the D$p$-brane
and each one of the 32 D9-branes of the background
(the $p$-$9 \oplus 9$-$p$ sector), in which tachyonic
modes could develop.

Let us first analyze the $p$-$p$ sector, whose
total vacuum amplitude is given by Eq. (\ref{ZopI}).
Performing the modular transformation to the open
string channel, and expanding the resulting expression
in powers of $q={\rm e}^{-\pi s}$, one can see \cite{gallot}
that 
\begin{equation}
{\cal Z}_{\rm open}^{p-p} \sim \int_0^{\infty} {ds \over 2s}\, 
s^{- {p+1 \over 2}}\,q^{-1}\,\Big[ \mu_p^2 - 2 
\,\mu_p \sin \big( {\pi \over 4}(9-p)\big) \Big] 
~~.
\label{asymptotic}
\end{equation}
The $q^{-1}$ behavior of the integrand signals the presence
of tachyons in the spectrum; 
therefore, in order not to have them, we must 
require that 
\begin{equation}
\mu_p = 2\,\sin \big(\frac{\pi}{4}(9-p)\big)~~.
\label{stability}
\end{equation}
Since $\mu_p$ has to be positive,
the only possible solutions are
\begin{equation}
 \begin{array}{|c|c|c|c|c|}\hline
 p     & -1 &    0     & \,7\, &      8        \\ \hline
 \mu_p &  2 & \sqrt{2} & 2 & \sqrt{2}      \\ \hline
 \end{array}
\label{tabular}
\end{equation}
{F}rom this table we see that in the type I theory there exist two
even non-BPS but stable D$p$-branes: 
the D-particle and the D8-brane.
Moreover, there exist also two 
odd non-BPS but stable D$p$-branes of type I: 
the D-instanton and the D7-brane.
Their tension is twice the one of the corresponding
type IIB BPS branes, in accordance with the fact that, as mentioned above, 
they can be simply interpreted as the superposition of a BPS brane
with an anti brane.
This classification of the stable non-BPS D-branes of type I based
on the table (\ref{tabular}) 
is in complete agreement with the results of Refs.~\cite{WITTEN}
derived from the K-theory of space-time.

Let us now analyze the $p$-$9 \oplus 9$-$p$ sector by considering the "mixed" 
cylinder amplitude (\ref{p-9}) which, after the modular transformation, reads
$$
{\cal Z}_{\rm open}^{9-p \oplus p-9}= 2^5 \mu_p V_{p+1} (8 \pi^2 \alpha')^{- {p+1 \over 2}}
 \int_0^{\infty}  {ds \over 2s} \,s^{- {p+1 \over 2}}
\left[ { f_3^{p-1}(q)f_2^{9-p} (q) \over f_1^{p-1}(q)f_4^{9-p}(q)}-
 { f_2^{p-1}(q)f_3^{9-p} (q) \over f_1^{p-1}(q)f_4^{9-p}(q)} \right]
$$
where the first and second term in the square brackets
account respectively for the 
NS and R sector, and the $f_i$'s are the standard one-loop 
functions \cite{pol}. 
This expression needs some comments. First, for $p=-1,0$
we see that there are no tachyons in the spectrum; moreover,
the values of $\mu_p$ for the D-instanton and D-particle 
are crucial in order to 
obtain a sensible partition function for open strings stretching between the 
non-BPS objects and the 32 D9-branes.
Secondly, for $p=7,8$ we directly
see the existence of a NS tachyon,  so 
that the corresponding branes are actually unstable \cite{gallot}. 
Hence, only the D-instanton and the D-particle are
fully stable configurations of type I string theory. 
Nevertheless, the strict relation connecting the D0-brane and the D(-1)-brane 
to the D8-brane and the D7-brane respectively, suggests that also the
latter may 
have some non trivial meaning.
Finally, we observe that the zero-modes of the Ramond sector
of these $p$-$9 \oplus 9$-$p$ strings
are responsible for the degeneracy of the non-BPS D$p$-branes
under the gauge group SO(32): in particular the
D-particle has the
degeneracy of the spinor representation of SO(32). 
Thus the D-particle accounts 
for the existence in type I of the non-perturbative non-BPS states 
required by the heterotic/type I duality.

We now present the basic
ideas and results about the gravitational
and gauge interactions of two stable non-BPS
D-particles of type I string theory (the detailed calculations and analysis
of these interactions can be found in \cite{gls}).
In the type I theory, D-branes interact 
via exchanges of both closed and open bulk strings. Since the dominant 
diagram for open strings has the topology of a disk, 
it gives a subleading (in the string coupling constant) contribution to 
the diffusion amplitude of two branes 
which is thus dominated by the cylinder diagram, {\it i.e.} by 
the exchange of closed strings. In the long distance limit, this
accounts for the gravitational interactions.
Let us now use this observation to calculate the 
dominant part of the scattering 
amplitude between two D-particles of type I moving with a relative
velocity $v$.
What we need to compute is the cylinder 
amplitude between the boundary state of a  static D-particle $\ket{D0}$
and the boundary state of a moving D-particle $\ket{D0,v}$ (the latter
is simply obtained by acting with a Lorentz boost on $\ket{D0}$ 
\cite{CANGEMI}).
Thus, the amplitude we are looking for is
${{\cal A}} = \langle D0 | P | D0,v \rangle +
\langle D0,v | P | D0 \rangle~$.
From this expression, we can extract the long range gravitational 
potential energy, which, in the non relativistic limit,
reads \cite{gls}
\begin{equation}
V^{\rm grav}(r) =  
(2 \kappa_{10})^2 \,{M_0^2 \over 7\, \Omega_8\, r^7}
\left(1 + \frac{1}{2} \,v^2 +o(v^2) \, \right)~~, 
\label{newton1}
\end{equation}
where $r$ is the radial coordinate, $\Omega_8$ is the 
area of the unit $8$-dimensional sphere,
$M_0=T_0/\kappa_{10}$ is the D-particle mass and $\kappa_{10}$
is the gravitational coupling constant in ten dimensions. 
Hence the boundary state calculation correctly reproduces
the gravitational potential we expect for a pair of D-particles 
in relative motion.

Although they are subdominant in the string coupling constant, 
the interactions of the D-particle with the open strings of the bulk
are nevertheless interesting because they account for the gauge interactions.
Since the non-BPS D-particles of type I are spinors of $SO(32)$, 
their gauge coupling 
is fixed by the spinorial representation 
they carry. The stringy description of such a coupling
has been provided in \cite{gls} where we have
shown that it is represented by an open string diagram with the 
topology of a disk with two boundary components, one lying
on the D9-branes from which the gauge boson is 
emitted, and the other lying on the D-particle.
At the points where the two boundary components join, we thus
have to insert a vertex operator $V_{90}$ (or $V_{09}$) that
induces the transition from Neumann to 
Dirichlet (or from Dirichlet to Neumann) boundary conditions in the nine 
space directions. 
As we have mentioned before, the $SO(32)$ 
degeneracy of the D-particle is due to the fermionic massless modes of the
open strings stretching between the D-particle and each of the 32 D9-branes;
therefore it is natural to think that
the boundary changing operators $V_{90}$
and $V_{09}$ are given by the vertex operators for these
massless fermionic modes. 
By construction, these operators carry Chan Paton factors
in the fundamental representation of SO(32), while the vertex 
operator $V_{\rm gauge}$ for the gauge boson carries a
Chan-Paton factor in the adjoint. 
As a consequence, this diagram must be considered as 
the one point function of the gauge boson in the background formed by a 
D-particle seen as an object in the bi-fundamental representation
of $SO(32)$. Hence, we do 
not see the entire gauge degeneracy of the D-particle because the 
degrees of freedom we use to describe it are not accurate enough. This is 
reminiscent from the fact that, in the boundary state formalism, also
the Lorentz 
degeneracy of a D-brane is hidden.
Using this result, we can easily compute the 
Coulomb potential energy $V^{\rm gauge}(r)$ for two
D-particles placed at a distance $r$. Indeed, this is
simply obtained by gluing two 1-point functions 
with a gauge boson propagator, and reads
\begin{equation}
{V}^{\rm gauge}(r) =
-\, \frac{g_{\rm YM}^2}{2}~\frac{1}{7\,\Omega_8\,r^7} 
\left(\delta^{AB}\,\delta^{CD}
-\delta^{AC}\,\delta^{DB}\right)~~,
\label{vgauge0}
\end{equation}
where $g_{\rm YM}$ is the gauge coupling constant in
ten dimensional type I string theory, and $A$, $B$, $C$ and  $D$
are indices in the fundamental representation of $SO(32)$.
We conclude by recalling that the non-BPS D-particles of type I are dual to
perturbative non-BPS states of the $SO(32)$ heterotic string which also 
have gravitational and gauge interactions among themselves. These can
be computed using standard perturbative methods and if one takes into
account the known duality relations and renormalization effects, one can 
explicitly check that they agree with the expressions (\ref{newton1})
and (\ref{vgauge0}). This agreement provides further dynamical
evidence of the heterotic/type I duality beyond the BPS level. 


\end{document}